\documentclass[aps,prd,nofootinbib,floatfix,amsmath,amssymb]{revtex4}
\usepackage{graphicx}
\begin{document}

\makeatletter
\newbox\slashbox \setbox\slashbox=\hbox{$/$}
\newbox\Slashbox \setbox\Slashbox=\hbox{\large$/$}
\def\pFMslash#1{\setbox\@tempboxa=\hbox{$#1$}
  \@tempdima=0.5\wd\slashbox \advance\@tempdima 0.5\wd\@tempboxa
  \copy\slashbox \kern-\@tempdima \box\@tempboxa}
\def\pFMSlash#1{\setbox\@tempboxa=\hbox{$#1$}
  \@tempdima=0.5\wd\Slashbox \advance\@tempdima 0.5\wd\@tempboxa
  \copy\Slashbox \kern-\@tempdima \box\@tempboxa}
\def\FMslash{\protect\pFMslash}
\def\FMSlash{\protect\pFMSlash}
\def\miss#1{\ifmmode{/\mkern-11mu #1}\else{${/\mkern-11mu #1}$}\fi}
\makeatother

\title{Implications of $D^0-\overline{D^0}$ on the rare top quark decays $t\to u\gamma$ and $t\to ug$}
\author{J. I. Aranda$^{(a)}$, A. Cordero--Cid$^{(b)}$, F. Ram\'\i rez--Zavaleta, J. J. Toscano$^{(c,d)}$, E. S. Tututi$^{(a)}$}
\address{$^{(a)}$Facultad de Ciencias F\'\i sico Matem\' aticas,
Universidad Michoacana de San Nicol\' as de
Hidalgo, Avenida Francisco J. M\' ujica S/N, 58060, Morelia, Michoac\'an, M\' exico. \\
$^{(b)}$ Facultad de Ciencias de la Electr\' onica, Benem\' erita Universidad Aut\' onoma de Puebla, Blvd. 18 Sur y Av. San Claudio, 72590, Puebla, Pue., M\' exico.\\
$^{(c)}$Facultad de Ciencias F\'{\i}sico Matem\'aticas,
Benem\'erita Universidad Aut\'onoma de Puebla, Apartado Postal
1152, Puebla, Puebla, M\'exico.\\
$^{(d)}$Instituto de F\'{\i}sica y Matem\' aticas, Universidad
Michoacana de San Nicol\' as de Hidalgo, Edificio C-3, Ciudad
Universitaria, C.P. 58040, Morelia, Michoac\' an, M\' exico.}

\begin{abstract}
The recently observed mass difference of the $D^0-\overline{D^0}$ mixing is used to predict the branching ratios of the rare top quark decays $t\to u\gamma$ and $t\to ug$ in a model independent way using the effective Lagrangian approach. It is found that $Br(t\to u\gamma)<4\times 10^{-4}$ and $Br(t\to ug)<2\times 10^{-3}$, which still may be within reach of the LHC collider.
\end{abstract}

\maketitle

\section{Introduction}	
With the recent start of the up--and--coming experimental program of the Large
Hadron Collider (LHC) at CERN, the search for new physics effects at the TeV
scale enters into an exciting era. In particular, due to the expected copious
production of top quark  events,   clues of physics  beyond the  Fermi scale
could be evidenced. The flavor changing neutral current (FCNC)
transitions of the top quark  are good processes to look for signals of
new physics, since the new dynamical effects are likely  more evident in
those top quark processes that are  forbidden or strongly suppressed in the
standard model (SM)~\cite{review}.  In the SM, the FCNC top transitions
$t\to cg$, $t\to c\gamma$, $t\to cZ$, and $t\to cH$ arise at the one--loop
level, being considerably GIM--suppressed, as  they have branching ratios
ranging from $10^{-10}$ to $10^{-13}$~\cite{DMPR,EHS1,EHS2,MP}.
The decays involving in the final state the quark $u$ instead of $c$ are
much more suppressed due to  Kobayashi--Maskawa effects. On the other hand,
the three  body $t\to cgg$~\cite{EFT} and $t\to c\bar{c}c$~\cite{CHTT1,CHTT2}
transitions have  also been studied within the context of the SM. It was
found~\cite{EFT,CHTT1,CHTT2} the  surprising result that the branching rations for
these decays are respectively  one order of magnitude larger and of the same order of magnitude
when compared with the two--body $t\to cg$  decay. Beyond the SM, FCNC top
quark decays have been the subject of considerable interest in the literature,
as they can have branching ratios much larger  than the SM ones. These
type of top transitions have been studied in some extended
theories such as the two--Higgs doublet model (THDM)~\cite{THDM11,THDM12,THDM13,THDM21,THDM22,THDM23,THDM24,THDM31,THDM32},
supersymmetric (SUSY) models with nonuniversal soft breaking~\cite{SUSY1,SUSY2,SUSY3,SUSY4,SUSY5,SUSY6,SUSY7,SUSY8,SUSY9,SUSY10,SUSY11},
SUSY models with broken R--parity~\cite{SUSYR1,SUSYR2}, models with extra $Z'$ gauge
 bosons~\cite{ZP}, and even more exotic scenarios~\cite{ES1,ES2,ES3,ES4}. Similar results
 were obtained within the context of effective theories~\cite{ETS1,ETS2,ETS3}.

In this work, we are interested in using the mass difference $\Delta M_D$
in the $D^0-\overline{D^0}$ mixing recently observed by the Babar~\cite{Babar}
and Belle~\cite{Belle} collaborations to impose a bound on the branching
ratios for  the FCNC top quark $t\to u\gamma$ and $t\to ug$ decays.
It turns out that due  to electromagnetic gauge invariance, the $t\to u\gamma$
transition is somewhat restricted, as it can occurs only  through two
electromagnetic gauge structures  of dipolar type, which are characterized
by Lorentz tensor structures of the form $\sigma_{\mu \nu}q^\nu$ (transition
magnetic dipole moment), with $q_\mu$  the photon  momentum,  or
$\gamma_5\sigma_{\mu \nu}q^\nu$ (transition electric dipole moment).
The $t\to ug$ decay occurs through the analogous chromomagnetic
and chromoelectric dipole moments. These interactions are naturally
suppressed,  as they can arise in renormalizable theories only through
quantum fluctuations  of one--loop or higher orders.

An outline of this paper is as follows. In section \ref{effective-lagrangian} we
briefly study the couplings arising from the  effective Lagrangian.
Section \ref{contributions} is devoted to obtain the leading contribution
of the $tu_i\gamma$ and $tu_ig$ couplings to the mass difference
$\Delta M_D$ of the $D^0-\overline{D^0}$ mixing. Finally,
in section \ref{final-remarks} we comment our results.

\section{Effective Lagrangian for the $tu_i\gamma$ and $tu_ig$ couplings}\label{effective-lagrangian}

The most general structure for the $tu\gamma$ and $tug$ couplings arise from the following dimension--six $SU_C(3)\times SU (2)_L\times U_Y(1)$--invariant effective Lagrangian:
\begin{eqnarray}
{\cal L}^{Up-FCNC}_{eff}&=&\frac{\alpha^{ij}_{uB}}{\Lambda^2}(\bar{Q}_i\sigma_{\mu \nu}B^{\mu \nu}u_j\tilde{\Phi})+
\frac{\alpha^{ij}_{uW}}{\Lambda^2}(\bar{Q}_i\sigma_{\mu \nu}W^{\mu \nu}u_j\tilde{\Phi})
\nonumber\\
&+&\frac{\alpha^{ij}_{uG}}{\Lambda^2}(\bar{Q}_i\sigma_{\mu \nu}G^{\mu \nu}u_j\tilde{\Phi}),
\end{eqnarray}
where $Q'$ and $u'$ stand for left-- and right--handed doublet and singlet of $SU_L(2)$, respectively and a sum over quark flavor is implied. Here, $B_{\mu \nu}$, $W_{\mu \nu}=\sigma^iW^i_{\mu \nu}/2$, and $G_{\mu \nu}^{a}=\lambda^aG_{\mu \nu}/2$ are the gauge tensors associated with the $U_Y(1)$, $SU_L(2)$, and $SU_C(3)$ groups, respectively. On the other hand, $\tilde{\Phi}=i\sigma^2 \Phi^*$, with $\Phi$ the SM Higgs doublet. In addition, the $\alpha^{ij}$ are the coefficients of general $3\times 3$ matrices defined in the flavor space, whereas $\Lambda$ represents the new physics scale. This parametrization allow us to treat in a model--independent way FCNC transitions in the up quark sector\footnote{FCNC transitions in the down quark sector are generated by the same type of operators with the replacements $u\to d$ and $\tilde{\Phi} \to \Phi$.} mediated by massless gauge bosons.

After spontaneous symmetry breaking, ${\cal L}^{Up-FCNC}_{eff}$ can be diagonalized as usual via the $V^u_L$ and $V^u_R$ unitary matrices, which relate gauge states to mass eigenstates. In particular, the $u_iu_j\gamma$ and $u_iu_jg$ vertices are given by the following Lagrangian:
\begin{eqnarray}
{\cal L}_{\overline{U}U\gamma (g)}&=&\frac{g}{2\sqrt{2}m_W}\Big(\overline{U}(\omega^{\gamma}P_R+\omega^{\gamma\dag}P_L)\sigma_{\mu \nu}UF^{\mu \nu}
\nonumber\\
&&
+
\overline{U}(\omega^{g}P_R+\omega^{g\dag}P_L)\sigma_{\mu \nu}\frac{\lambda^a}{2}UG^{a\mu \nu}\Big),
\end{eqnarray}
where $\overline{U}=(\bar{u},\bar{c},\bar{t})$ are vectors in the flavor space. Notice the presence of the global factor $g/\sqrt{2}m_W$, which arises from the Higgs mechanism and is common to both electromagnetic and strong couplings. In the above expression, $\omega^{\gamma}$ and $\omega^g$ are flavor matrices given by
\begin{eqnarray}
\omega^{\gamma}&=&\Big(\frac{v}{\Lambda}\Big)^2V^u_L\big(c_W\alpha_{uB}+\frac{1}{2}s_W\alpha_{uW}\big)V^{u\dag}_R,\\
\omega^g&=&\Big(\frac{v}{\Lambda}\Big)^2V^u_L\alpha_{uG}V^{u\dag}_R,
\end{eqnarray}
where $v\approx 246$ GeV is the Fermi scale and $c_W$ ($s_W$) stands for cosine (sine) of the weak angle. To generate vector--mediated FCNC effects at the tree level in the effective theory, it is assumed that the $V^u_{L,R}$ matrices diagonalize the Yukawa mass matrix $Y^u$ but not the matrices $\alpha_{uB(uW)}$ and $\alpha_{uG}$. From the above expressions, it is immediate to derive the vertex functions for the $tu_i\gamma$ and $tu_ig$ ($u_i=u,c$) couplings, which can be written as
\begin{equation}
\Gamma^{(\gamma,g)}_{\mu}(q)=\frac{g}{\sqrt{2}m_W}\Big(\kappa^{(\gamma,g)}_{tu_i}+i\tilde{\kappa}^{(\gamma,g)}_{tu_i}\gamma_5\Big)\sigma_{\mu \nu}q^\nu \delta^{(\gamma,g)},
\end{equation}
where $\delta^\gamma=1$ and $\delta^g=\lambda^a/2$. In addition, $\kappa^{(\gamma,g)}_{tu_i}=Re(\omega^{(\gamma,g)}_{tu_i})$ and $\tilde{\kappa}^{(\gamma,g)}_{tu_i}=Im(\omega^{(\gamma,g)}_{tu_i})$. As we will see below, it is not possible to isolate the contribution of the $tuV$ and $tcV$ ($V=\gamma,g$) couplings from the $D^0-\overline{D^0}$ mixing. This contribution is given through an amplitude that depends symmetrically on both vertices. Due to this, it is not possible to obtain bounds for the $t\to c\gamma$ and $t\to cg$ decays without making additional assumptions.

\section{$tu_i\gamma$ and $ tu_ig$ contributions to $D^0-\overline{D^0}$ mixing}\label{contributions}

\begin{figure}
\centering
\includegraphics[width=3.5in]{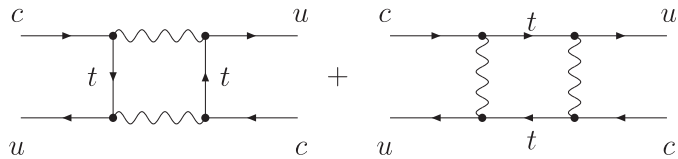}
\caption{\label{FIG1} Simultaneous contribution of the $tc\gamma$ and $tcu$ couplings to the $D^0-\overline{D^0}$ mixing. The contribution of the $tcg$ and $tug$ couplings also is given through this type of diagrams.}
\end{figure}

We now are in position of deriving the contribution of the $tu_i\gamma$ and $tu_ig$ vertices to the mass difference $\Delta M_D$ of the $D^0-\overline{D^0}$ mixing. These contributions occur through short distance effects characterized by the box diagrams shown in FIG.\ref{FIG1}. Neglecting the external momenta, the $tu_i\gamma$ and $tu_ig$ couplings generate an amplitude given by:
\begin{eqnarray}
T&=&2\Bigg(\frac{g}{\sqrt{2}m_W}\Bigg)^4\Bigg(|\omega^\gamma_{tu}|^2|\omega^\gamma_{tc}|^2
+\frac{11}{18}|\omega^g_{tu}|^2|\omega^g_{tc}|^2\Bigg)
\nonumber\\
&&\times\int \frac{d^Dk}{(2\pi)^D}
\frac{
(\bar{u}_L\sigma_{\mu \lambda}k^\lambda \pFMSlash{k}\sigma_{\alpha \rho}k^\rho c_L)(\bar{u}_L\sigma_{\mu \eta}k^\eta\pFMSlash{k}\sigma_{\mu \xi}k^\xi c_L)}{(k^2)^2(k^2-m^2_t)^2}.
\label{qdiv}
\end{eqnarray}
The integral involved in (\ref{qdiv}) is of the  form
\begin{displaymath}
I^{(a,b)}_{\alpha\beta\mu\nu\lambda\rho}\equiv\int\frac{d^Dk}{\left(2\pi\right)^D}
\frac{k_{\alpha}k_{\beta}k_{\mu}k_{\nu}k_{\lambda}k_{\rho}}{(k^2)^a(k^2-m^2_t)^b},
\end{displaymath}
with $a=2$ and $b=2$, and it diverges quadratically at $D=4$. As discussed in reference~\cite{BL}, the use of cutoffs to regularize quadratic or higher divergent integrals would violate the decoupling theorem, as this scheme introduce a strong dependence on the new physics scale. In our case, the use of a cutoff leads to an amplitude proportional to $\Lambda^2$, which would overestimate our prediction. Only a logarithmic dependence on the new physics scale is physically acceptable~\cite{BL,W} (for some applications of dimensional regularization in  effective Lagrangians see, for instance,  Refs.~\cite{Tosca1,Tosca2,Tosca3}). Thus, we do not use a cutoff to regularize our amplitude, instead we will use dimensional regularization, which allow one to convert the quadratic divergence into a logarithmic one through the following 
limit~\cite{delbourgo-scadron1,delbourgo-scadron2}:
\begin{eqnarray}
\lim_{\varepsilon\to 0}\left(I^{(2,2)}_{\alpha\beta\mu\nu\lambda\rho}-
2m^{2}_{t}I^{(2,3)}_{\alpha\beta\mu\nu\lambda\rho}\right)&=&
-\frac{im^2_t}{3072\pi^2}\bigg[g_{\alpha\beta}\left(g_{\mu\nu}g_{\lambda\rho}
+g_{\mu\lambda}g_{\nu\rho}+g_{\nu\lambda}g_{\mu\rho}\right)
\nonumber\\
&&+ g_{\alpha\mu}\left(g_{\beta\nu}g_{\lambda\rho}+g_{\beta\lambda}g_{\nu\rho}+
g_{\beta\rho}g_{\nu\lambda}\right)
\nonumber\\
&&+g_{\alpha\nu}\left(g_{\beta\mu}g_{\lambda\rho}+g_{\beta\lambda}g_{\mu\rho}+
g_{\beta\rho}g_{\nu\lambda}\right)
\nonumber\\
&&+g_{\alpha\lambda}\left(g_{\beta\rho}g_{\mu\nu}+g_{\beta\nu}g_{\mu\rho}+
g_{\beta\mu}g_{\nu\rho}\right)
\nonumber\\
&&+g_{\alpha\rho}\left(g_{\beta\lambda}g_{\mu\nu}+g_{\beta\mu}g_{\nu\lambda}+
g_{\beta\nu}g_{\mu\lambda}\right)\bigg].
\label{eq2}
\end{eqnarray}
The arising integral with logarithmic divergence,
$I^{(2,3)}_{\alpha\beta\mu\nu\lambda\rho}$,  can be treated in the usual
manner, hence a renormalization scheme can directly be applied  to the
amplitude in (\ref{qdiv}).

Once the integral is solved, one obtains the result for the amplitude:
\begin{eqnarray}
T &=&-\frac{i\alpha^2}{s^4_Wm^2_W}\Big(\frac{21}{4}\Big)\Big(\frac{m_t}{m_W}\Big)^2\Bigg(|\omega^\gamma_{tu}|^2|\omega^\gamma_{tc}|^2
+\frac{11}{18}|\omega^g_{tu}|^2|\omega^g_{tc}|^2\Bigg)
\nonumber\\
&&\times(\bar{u}_L\gamma_\mu c_L)(\bar{u}_L\gamma^\mu c_L)\Bigg(\Delta+\log\Big(\
\frac{\mu^2}{m^2_t}\Big)+\frac{19}{12}\Bigg),
\end{eqnarray}
where the ultraviolet divergence is contained in $\Delta=\frac{2}{\varepsilon}-\gamma_E+\log(4\pi)$, with $\varepsilon=4-D$ and $\gamma_E$ the Euler's constant. Here, $\mu$ is the dimensional regularization scale. Following Refs.\cite{BL,W,Tosca1,Tosca2,Tosca3,RMS1,RMS2}, the divergent
term in this amplitude can be absorbed by renormalizing the appropriated coefficients of the complete effective Lagrangian since it already contains all the
invariants allowed by the SM symmetry. The invariants needed to absorb the divergences are of the form $(\alpha_{12}/\Lambda^2)(\bar{q}_1\gamma_\mu q_2)(\bar{q}_1\gamma^\mu q_2)$, with $\bar{q}_1=(\bar{u}_L,\bar{d}_L)$ and $\bar{q}_2=(\bar{c}_L,\bar{s}_L)$, etc. Using the $\overline{MS}$ renormalization scheme with $\mu=\Lambda$, the renormalized amplitude can be written as follows:
\begin{eqnarray}
T &=&-\frac{i\alpha^2}{s^4_Wm^2_W}\Big(\frac{21}{4}\Big)\Big(\frac{m_t}{m_W}\Big)^2\Bigg(|\omega^\gamma_{tu}|^2|\omega^\gamma_{tc}|^2
+\frac{11}{18}|\omega^g_{tu}|^2|\omega^g_{tc}|^2\Bigg)
\nonumber\\
&&\qquad \times(\bar{u}_L\gamma_\mu c_L)(\bar{u}_L\gamma^\mu c_L)f(\Lambda/m_t),
\end{eqnarray}
where we have defined the dimensionless function $f(\Lambda/m_t)=
\log\Big(\frac{\Lambda^2}{m^2_t}\Big)+\frac{19}{12}$. It is easy to see that this amplitude can be obtained directly from a four--quark effective interaction
\begin{equation}
{\cal L}_{eff}=-\frac{\alpha^2}{s^4_Wm^2_W}\Big(\frac{21}{16}\Big)\Big(\frac{m_t}{m_W}\Big)^2\Bigg(|\omega^\gamma_{tu}|^2|\omega^\gamma_{tc}|^2
+\frac{11}{18}|\omega^g_{tu}|^2|\omega^g_{tc}|^2\Bigg)f(\Lambda/m_t)Q_1,
\end{equation}
where the dimension--six operator $Q_1=(\bar{u}_L\gamma_\mu c_L)(\bar{u}_L\gamma^\mu c_L)$ is well known in the literature~\cite{Operators}. In the above expression, a factor of $1/4$ was introduced in order to compensate the $(2!)(2!)$ Wick contractions.

On the other hand, the  mass difference $D$ is given by
\begin{equation}
\Delta M_D=\frac{1}{M_D}Re<\overline{D^0}|{\cal H}_{eff}=-{\cal L}_{eff}|D^0>,
\end{equation}
which in our case takes the form
\begin{eqnarray}
\Delta M_D&=& \frac{\alpha^2}{s^4_Wm^2_W}\Big(\frac{21}{16}\Big)\Big(\frac{m_t}{m_W}\Big)^2\Bigg(|\omega^\gamma_{tu}|^2|\omega^\gamma_{tc}|^2
+
\frac{11}{18}|\omega^g_{tu}|^2|\omega^g_{tc}|^2\Bigg)f(\Lambda/m_t)<Q_1>
\nonumber \\
&=&\Big(\frac{21}{24}\Big)\frac{\alpha^2}{s^4_W}\Big(\frac{m_t}{m_W}\Big)^2\Big(\frac{f_D}{m_W}\Big)^2M_D\Bigg(|\omega^\gamma_{tu}|^2|\omega^\gamma_{tc}|^2
+
\frac{11}{18}|\omega^g_{tu}|^2|\omega^g_{tc}|^2\Bigg)f(\Lambda/m_t)B_1,
\nonumber\\
\end{eqnarray}
where the expressions for $<Q_1>$ given in Ref.\cite{Operators} were used. Here, $f_D$ is the $D$ decay constant. We will use the CLEO Collaboration determination $f_D=222.6\pm 16.7 MeV $~\cite{CLEO}. The factor $B_1$ is unknown, but lattice calculation~\cite{Lattice} leads to $B_1\approx 0.8$, although in vacuum saturation and in the heavy quark limit it approximates to the unity. In our numerical analysis, we will use $B_1=1$. Using the value $M_D=1.8646$ GeV, one obtains
\begin{equation}
\Delta M_D=6.28\times 10^{-8}f(\Lambda/m_t)\Bigg(|\omega^\gamma_{tu}|^2|\omega^\gamma_{tc}|^2
+\frac{11}{18}|\omega^g_{tu}|^2|\omega^g_{tc}|^2\Bigg).
\end{equation}
From the experimental side, the Heavy Flavor Averaging Group~\cite{HFAG} interpretation of the current data leads to a mass difference of $D^0-\overline{D^0}$ given by $\Delta M_D=(1.4\pm 0.5)\times 10^{-15}$ GeV \cite{PDG}. To bound the $\kappa_i$ coefficients, we demand that the above contribution does not exceed the experimental uncertainty, which leads to
\begin{equation}
|\omega^\gamma_{tu}|^2|\omega^\gamma_{tc}|^2
+\frac{11}{18}|\omega^g_{tu}|^2|\omega^g_{tc}|<7.96\times 10^{-9},
\end{equation}
where it was assumed that $f(\Lambda/m_t)\sim O(1)$. This assumption is reasonable~\cite{MQ}, as a new physics scale in the TeVs region is expected. Considering the electromagnetic and strong contributions one at the time, one obtains
\begin{eqnarray}
\label{b1}
|\omega^\gamma_{tu}||\omega^\gamma_{tc}|&<&8.92\times 10^{-5}, \\
\label{b2}
|\omega^g_{tu}||\omega^g_{tc}|&<&1.14\times 10^{-4}.
\end{eqnarray}
From general considerations, one can expect that $\omega^{\gamma,g}_{tc}>\omega^{\gamma,g}_{tu}$, so that a conservative point of view allow us to obtain the bounds
\begin{eqnarray}
|\omega^\gamma_{tu}|^2&<&8.92\times10^{-5}, \\
|\omega^g_{tu}|^2&<&1.14\times10^{-4}.
\end{eqnarray}
It should be noticed that it is not possible to bound the $tc\gamma$ and $tcg$ without making additional assumptions.

On the other hand, the branching ratios for the $t\to u\gamma$ and $t\to ug$ decays are given by
\begin{eqnarray}
Br(t\to u\gamma)&=&\frac{\alpha}{4s^2_W}\Big(\frac{m_t}{\Gamma_t}\Big)\Big(\frac{m_t}{m_W}\Big)^2|\omega^\gamma_{tu}|^2, \\
Br(t\to ug)&=&\frac{\alpha}{s^2_W}\Big(\frac{m_t}{\Gamma_t}\Big)\Big(\frac{m_t}{m_W}\Big)^2|\omega^g_{tu}|^2.
\end{eqnarray}
From these expressions and from Eqs.(\ref{b1},\ref{b2}), we obtain the following bounds for these decays
\begin{eqnarray}
Br(t\to u\gamma)&<&3.9\times 10^{-4}, \\
Br(t\to ug)&<&2.02\times 10^{-3},
\end{eqnarray}
where the approximation $\Gamma_t\approx \Gamma(t\to bW)=1.55$ GeV for the total top decay width was used.

It is worth comparing our results with those predictions obtained in some specific models. Most of the known results are on the $t\to c\gamma$ and $t\to cg$ decays, which are strongly suppressed in the SM, with branching ratios of order of $10^{-13}$ and $10^{-11}$~\cite{EHS1,EHS2}, respectively. The branching ratios for the decays involving the quark $u$ instead of $c$ are even more suppressed by a Kobayashi--Maskawa factor of $(V_{ub}/V_{cb})^2\approx 10^{-2}$. Beyond the SM, these branching ratios are considerably enhanced. For instance, in the THDM-III, the $t\to c\gamma$ and $t\to cg$ decays can have a branching ratio in the range $10^{-7}-10^{-12}$ and $10^{-4}-10^{-8}$~\cite{THDM21,THDM22,THDM23,THDM24}, respectively. On the other hand, the respective predictions in some SUSY models~\cite{SUSY1,SUSY2,SUSY3,SUSY4,SUSY5,SUSY6,SUSY7,SUSY8,SUSY9,SUSY10,SUSY11} are of order of $10^{-5}$ and $10^{-3}$. An effective Lagrangian analysis of Higgs--mediated FCNC leads to branching ratios for these decays of order of $10^{-7}$ and $10^{-6}$~\cite{ZP}. All these predictions are consistent with our bound, as it is expected that the branching ratios for the $t\to u\gamma$ and $t\to ug$ decays  are lower by  two or higher orders of magnitude than those involving the quark $c$.

\section{Final remarks}\label{final-remarks}
To conclude, we would like to comment on the possible detection of the $t\to u\gamma$ and $t\to ug$ decays at the LHC. On the light of our bounds for their branching ratios, one can conclude that they are within reach of this collider, given the important production of $t\bar{t}$ events of several millions per year. In a purely statistical basis those channels with branching ratios larger than about $10^{-6}$ do have the chance of being detected. However, the observability of a particular channel decay depends on several factors, background and systematics may reduce this value by several orders of magnitude depending of the particular signature. For instance, the $t\to ug$ decay would require a large branching in order to be detected as it is swamped by hadronic backgrounds. However, the important difference of about three orders of magnitude between the observability parameter and our bound makes quite possible the observation of this channel. As far as the $t\to u\gamma$ mode is concerned, it could be detected even with a relatively small branching ratio because it would be produced in a cleaner environment.  In conclusion, our analysis suggests that the recent observation of $D^0-\overline{D^0}$ mixing does not exclude the possibility of observing the rare top quark decays $t\to u\gamma$ and $t\to ug$ at the LHC. Since the $tc\gamma$ and $tcg$ couplings are not directly constrained by this experimental result, their observation at the LHC would be more probable.

\section*{Acknowledgments}
We acknowledge financial support from CONACYT and
SNI (M\' exico).


\end{document}